# A Weakest Chain Approach to Assessing the Overall Effectiveness of the 802.11 Wireless Network Security

Berker Tasoluk[1] and Zuhal Tanrikulu[2]

[1]Department of Informatics, Istanbul University, Istanbul, Turkey
`berker@berker.cc`

[2]Department of Management Information Systems, Bogazici University, Istanbul, Turkey
`tanrikul@boun.edu.tr`

## ABSTRACT

*This study aims to assess wireless network security holistically and attempts to determine the weakest link among the parts that comprise the 'secure' aspect of the wireless networks: security protocols, wireless technologies and user habits. The assessment of security protocols is done by determining the time taken to break a specific protocol's encryption key, or to pass an access control by using brute force attack techniques. Passphrase strengths as well as encryption key strengths ranging from 40 to 256 bits are evaluated. Different scenarios are planned and created for passphrase generation, using different character sets and different number of characters. Then each scenario is evaluated based on the time taken to break that passphrase. At the end of the study, it is determined that the choice of the passphrase is the weakest part of the entire 802.11 wireless security system.*

## 1. INTRODUCTION

The 802.11 wireless Internet and network usage is increasing daily for the benefits and ease of use offered to its users. Almost everywhere that people go, including airports, cafes, their homes and workplaces, and hotels, they use wireless networks. This increasing usage also brings some security concerns. Users require that their sensitive information not be seen or altered by anyone else. In addition, users require that any security measure taken will not interfere with their working habits; they want the security to be transparent to their normal network usage.

Many wireless network protocols exist today. Some of them were used in networks a few years ago and are no longer used, and some of them are still used in today's networks. The standards are historically outlined below.

802.11: This is the first standard published by IEEE in 1997. It operates on a 2.4 Ghz frequency band and can increase up to 2 Mbps. It is not used at the moment, since the 2 Mbps speed does not meet users' needs.

802.11a: This is the second standard published in 1999. It operates on a 5 Ghz band and can operate at speeds up to 54 Mbps. The main advantage of 802.11a is that it operates at a 5 Ghz frequency band, which makes it less prone to interference. It also has more non-overlapping channels, which also reduces the possibility of interference. The main drawback of this protocol is that it is incompatible with other wireless protocols such as 802.11b and 802.11g. It is seldom used at the moment [1].

802.11b: Before the 802.11g protocol, this was the protocol that was most commonly used. It operates on a crowded 2.4 Ghz frequency band and can reach speeds up to 11 Mbps.

802.11g: This protocol is similar to the 802.11b, as it operates on a 2.4 Ghz band, but with speeds of up to 54 Mbps. It is compatible with 802.11b, and is the most widely used protocol for new installations.

802.11n: It is designed to produce speeds up to 600 Mbps. This standard operates at 2.4 Ghz and/or 5 Ghz. At the moment, some 802.11n compatible laptops and wireless access points can be seen in the market [2].

There are numerous wireless security protocols that can be deployed in wireless networks. Wireless network security can be divided into three basic layers: [3] Wireless LAN Layer, Access Control Layer and Authentication Layer.

Wireless LAN Layer Security Protocols is responsible for the encryption and decryption of data. These protocols are as follows:

WEP (Wired Equivalent Privacy): Despite its name, this is the weakest form of encryption security for wireless networks. WEP can use a 64-bit or 128-bit stream RC4 cipher with a 24-bit initialization vector (IV), which leads to an effective key size of 40 or 104 bits. RC4 is also a weak cipher that can be easily cracked within minutes or hours.

WPA (Wi-Fi Protected Access): WPA still uses RC4 algorithm with doubled initialization vector (IV) and supports key lengths up to 152 bits. Recent attacks have targeted the WPA algorithm and achieved success on certain specific WPA implementations [4].

WPA2: WPA2 uses AES algorithm with varying lengths of 128, 192 and 256 bits. The AES algorithm is considered more secure than RC4 and uses block cipher, which involves 128 bits of data [5].

Access Control Layer Security Protocols is responsible for allowing only legitimate users access to network resources. For this to happen, this layer of protocols talks to authentication layer protocols. The protocols are outlined below.

802.1X: Originally developed for wired network access control, it is now also widely deployed for wireless networks. Many institutions integrate dot1X into their existing local area networks and wireless networks. LAN switches and wireless access points must support dot1X in order to use this protocol. 802.1X simply obtains user credentials and asks an authentication server whether to allow this user or not. It typically asks a RADIUS server, although it is not mandatory to use RADIUS.

RADIUS (Remote Access Dial-In User Service): As its name suggests, RADIUS was originally designed for dial-in user access, but it can be used for any access control purpose, including dot1X integration. It is a central database of usernames and passwords, as well as some user attributes such as VLAN ID [6].

EAP (Extensible Authentication Protocol): This is an intermediary protocol between the access control layer and authentication layer security protocols. It can work with a variety of authentication protocols; future authentication protocol support is embedded into the protocol (the word 'extensible' is used for this behavior).

Authentication Layer Security Protocols responds to access control layer security protocols. It decides who can join the network, and the decision is implemented by access control layer security protocols. These protocols are as follows:

TLS: This is the IEEE standardized version of Netscape's well-known, widely used SSL protocol. It relies on using certificates and symmetric/asymmetric ciphers. The Wi-Fi alliance required vendors to support at least the TLS protocol. Therefore, it is widely deployed with WPA, and the interoperability between different vendors is high [7]. For wireless security, TLS is generally used more often than EAP.

Kerberos: Invented at MIT during the 1980s, this protocol has undergone some changes, been standardized by IETF and reached its fifth version. It relies on the ticket granting concept and symmetric ciphers, and is still used in some deployments.

LEAP (Cisco Light EAP): This is a Cisco devised protocol that is well suited to the 802.1X scheme. It is also based on MS-CHAPv1, which is considered an unsecure protocol.

PEAP (Protected EAP): The PEAP's main goal is to bring confidentiality to an EAP session and safeguard the credentials. If a secure protocol like TLS is not used within EAP, it is strongly recommended to use PEAP instead of EAP.

Other than these protocols, there is another component of the entire system: The users. Often, weak passwords are used by them; either short number of characters or dictionary words are used. In a work performed by Klein, even using a small 62,727 word dictionary; 25% of 13,797 passwords were cracked [8].

## 2. PROBLEM STATEMENT

The afore mentioned wireless technologies and protocols provide different levels of security to users. In a typical wireless network environment, a blend of these protocols is used. Although some protocols and security mechanisms provide adequate security, others may not. Using a strong security mechanism with a weak one may decrease the overall security posture of the wireless network. It may also provide its users with a false sense of safety. Most of the time, only a few aspects of the wireless network security are considered, while other aspects are omitted. Is it possible to analyze all pieces and components of the entire system? Do people realize that they are only as secure as the weakest component of the entire system? This study aims to find answers to these problems.

## 3. OBJECTIVES

This study aims to explore the 802.11 wireless network security protocols; identifying the underlying ciphering algorithms' strengths based on used key's bit length, and identifying the used passphrases' strength based on the number and type of passphrase characters used. Then, the weakest chain among those two components will be determined and a best practice approach to increase the overall security of a wireless network system will be suggested. This is the main objective of the study.

## 4. METHOD

The assessment of security protocols is done by determining the time taken to break a specific protocol's encryption key, or to pass an access control. This time is directly proportional to all possible keys that an encryption algorithm may have, or all possible values that an authentication protocol may require from the user. It is assumed that application specific integrated circuits (ASIC) are used to break any given encryption key.

In this study, the brute force resistance of security protocols is evaluated by assuming a 10,000 ASIC chip setup is established, and $1 \times 10^{12}$ keys per second are tried. By using 10,000 ASIC circuits, 100M x 10,000 = $1 \times 10^{12}$ keys per second can be tried. A 100 MHz ASIC can try 100M keys per second [9]. Schäfer [10] stated that it is feasible to achieve $10^6$ encryptions / μ second in today's technology, which is equivalent to $1 \times 10^{12}$ keys per second.

Other cryptanalysis methods are not considered, as they are beyond the scope of this study.

## 5. RESULTS

The assessment of a wireless security protocol can be found below by investigating the time required to break that specific protocol.

Table 1: WEP / WPA / WPA2 Key Strength: Cipher Strength

| Key Length | Typically Found in | Duration of Cracking |
|---|---|---|
| 40-bits | WEP | $2^{40} / 10^{12}$ = **1.09951163 seconds** |
| 104-bits | WEP | $2^{104} / 10^{12}$ = **6.42724504 × $10^{11}$ years** |
| 256-bits | WPA / WPA2 | $2^{256} / 10^{12}$ = **3.67174306 × $10^{57}$ years** |

The above table reveals that, without considering the cryptanalysis methods, breaking 40-bit keys takes just a second, but breaking 104-bit and 256-bit keys requires a considerable amount of time. Here, the weaknesses of the WEP algorithm and other attack methods may further reduce the key-breaking times.

Theoretically, ten totally random hexadecimal characters as a passphrase for 40-bit WEP, 26 totally random hexadecimal characters as a passphrase for 104-bit WEP, and 64 totally random hexadecimal characters as a passphrase for WPA/WPA2 Phrase Shared Key Mode can be used. However, in practice, it is known that this not the case. In their study, Morris and Thompson [11] found that among 3,289 user passwords investigated, eighty-six percent of all passwords were equal to or less than six characters long or were easily found in a dictionary or name list.

WPA2 passphrases are another kind of 'user password' and have the same vulnerabilities as traditional user passwords. A brute force attack analysis of the mentioned passphrases is outlined in Table 2.

In the table, the duration of crack times is calculated using two variables: the first variable being the encryption algorithm's passphrase strength (second column of the table), and the second variable being the character sets chosen to generate the passphrase (first line of the table).

With a 256-bit encryption algorithm, there are $2^{256}$ different key possibilities, and this leads to $3.67174306 \times 10^{57}$ years of cracking time (Table 1). When using 63-character passphrases, which are created using 26-character sets or higher, even higher calculated values are possible. These values have no practical significance and are reduced to $3.67174306 \times 10^{57}$ years on the last line of Table 2.

Table 2: WEP / WPA / WPA2 Key Strength: User Given Passphrase Strength

| Encryption Algorithms' Key & Passphrase Lengths | | Passphrase Character Sets | | | | |
|---|---|---|---|---|---|---|
| Encryption Key Length (Bits) | Passphrase Length (Characters) | Using 10 character set | Using 26 character set | Using 36 character set | Using 52 character set | Using 62 character set |
| 40-bits WEP | 5 Characters | 100 nano-seconds | 11.881376 micro-seconds | 60.466176 micro-seconds | 380.204032 micro-seconds | 916.132832 micro-seconds |
| 256-bits WPA/WPA2 PSK | 8 Characters | 100 micro-seconds | 0.208827065 seconds | 2.82110991 seconds | 53.4597285 seconds | 3.63900176 minutes |
| 104-bits WEP | 13 Characters | 10 seconds | 28.7170471 days | 5.40552424 years | 644.093292 years | 6 338.6573 years |
| 256-bits WPA/WPA2 PSK | 16 Characters | 2.77777778 hours | 1 381.90719 years | 252 200.139 years | 90 564 669.6 years | $1.51067952 \times 10^{9}$ years |
| 256-bits WPA/WPA2 PSK | 32 Characters | $3.16887646 \times 10^{12}$ years | $6.02632354 \times 10^{25}$ years | $2.00717544 \times 10^{30}$ years | $2.58828625 \times 10^{35}$ years | $7.20177208 \times 10^{37}$ years |
| 256-bits WPA/WPA2 PSK | 63 Characters | $3.16887646 \times 10^{43}$ years | $4.40784416 \times 10^{69}$ years (calculated) | $3.53152965 \times 10^{78}$ years (calculated) | $4.06551865 \times 10^{88}$ years (calculated) | $2.63986518 \times 10^{93}$ years (calculated) |
| 256-bits WPA/WPA2 PSK | 63 Characters | $3.16887646 \times 10^{43}$ years | $3.67174306 \times 10^{57}$ years (practical) | $3.67174306 \times 10^{57}$ years (practical) | $3.67174306 \times 10^{57}$ years (practical) | $3.67174306 \times 10^{57}$ years (practical) |

Considering the duration of crack times, it is assumed that using a combination of key strength and a chosen character set is safe if it takes longer than 89.78 years to decipher the key; this is the greatest "average life expectancy of a human at birth" among all the countries [12].

Here, it is assumed that the following character sets are used when producing passphrases. Usage of different character sets yields different levels of security.

- Only Arabic numbers:

  0 1 2 3 4 5 6 7 8 9 (10 characters)

- Only Latin lowercase letters:

  a b c d e f g h i j k l m n o p q r s t u v w x y z (26 characters)

- Only Latin uppercase letters:

  A B C D E F G H I J K L M N O P Q R S T U V W X Y Z (26 characters)

- Either lowercase or uppercase letters + numbers (36 characters)

- Both lowercase and uppercase letters (52 characters)

- Both lowercase and uppercase letters + numbers (alphanumeric set) (62 characters)

From the table, it can be concluded that using five characters (typical for 40-bit WEP) or eight characters (minimum character number required for WPA/WPA2 PSK mode) offers no security. Even when producing the keys completely randomly from the alphanumeric set, it can be cracked in a very short time frame.

Generating 13 character passphrases by using only numbers or only lowercase or uppercase letters offers poor security. Only using numbers and letters combined seems to provide adequate security at 13 character passphrases, given that the passphrase is randomly produced.

Using a 16-character scenario seems to be secure unless the passphrase uses only numbers.

Using 32 or 63 character seems safe even if the passphrase only contains numbers.

Table 3: Minimum Number of Characters in the Set to Ensure a 'Secure' Passphrase

| Encryption Key Length (Bits) | Passphrase Length (Characters) | Minimum Number of Characters in the Set |
|---|---|---|
| 40-bits WEP | 5 Characters | **19 516.289** |
| 256-bits WPA/WPA2 PSK | 8 Characters | **480.284174** |
| 104-bits WEP | 13 Characters | **44.6840764** |
| 256-bits WPA/WPA2 PSK | 16 Characters | **21.9153867** |
| 256-bits WPA/WPA2 PSK | 32 Characters | **4.68138726** |
| 256-bits WPA/WPA2 PSK | 63 Characters | **2.19032075** |

For each of the passphrase lengths, Table 3 gives the minimum number of character sets that must be used in order to ensure the key can be recovered in 89.78 years. These values are calculated by using the following formula:

$$x = \sqrt[y]{\left(89.78 \, years * 365 \frac{days}{years} * 86400 \frac{seconds}{days} * 10^{12} \frac{characters}{seconds}\right)}$$

where y denotes the passphrase length and x denotes the minimum number of characters in the set.

## 6. CONCLUSIONS

From these findings, it can be concluded that even if WPA2 with a 256-bit AES algorithm is used and this algorithm's code breaking requires 3.67 x $10^{57}$ years, using weak passphrases may reduce this security level and code breaking times to days or even minutes. The choice of the passphrase seems to be the weakest part of the entire 802.11 wireless security system. It is advised to ensure that adequate procedures are in place that guarantee that chosen passphrases are generated completely randomly, and that passphrases include the minimum required characters and are generated from character sets containing the minimum number of elements.

The assessment is completed by assuming that 10,000 x 100 MHz ASIC circuits are used for deciphering the key. Using the CPUs of personal computers or distributed computing may be other alternatives.

This work only focuses on brute force attacks. Future studies may focus on dictionary attacks, or using word lists. Furthermore, the entropy of language may be considered. For example, Shannon estimates that the English language has entropy of approximately 2.3 bits per letter [13], which means that even using a non-dictionary word or phrase may yield a less secure key than normally expected.

This work does not take into account wireless security protocol weaknesses. The key length of any given protocol is assumed to be the ultimate criterion. Future studies may focus on protocol weaknesses and attacks targeted on them.

**Authors**

**Berker Tasoluk** received his B.S. in Electrical and Electronics Engineering from the Middle East Technical University in Ankara, and his M.S. degree from the Informatics Institute at the same university. He started his career as a part-time researcher at "The Scientific and Technological Research Council of Turkey (TUBITAK – Bilten)", Computer Networks Groups. Then he worked on many areas of information technology as a network engineer, security engineer, product manager, project manager, IT auditor and consultant. Currently he is pursuing his studies on Informatics at Istanbul University and working as a network and security consultant at a leading IT integrator.

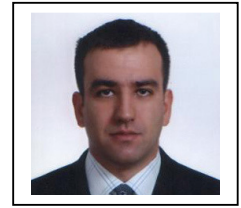

**Zuhal Tanrikulu** is an Associate Professor of Management Information Systems at Bogazici University in Istanbul. She received her B.S. in Electronics and Communication Engineering, M.S. in Computer Engineering, and Ph.D. in Management and Organization, specializing in MIS. Her current research interests focus on the implementation of information technology in organizations, analysis and design of integrated information systems, information systems security and management, web service-based information systems, and learning management systems. She is active in teaching algorithms, programming, and information systems management courses. She has been published in Educause Quarterly, International Review on Computers and Software, Yonetim (Management), and the Journal of the Faculty of Education.

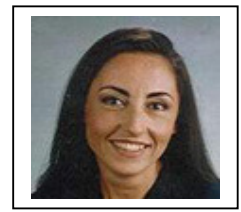